\documentclass[a4paper,twocolumn,superscriptaddress,11pt]{quantumarticle}
\pdfoutput=1
\makeatletter
\providecommand{\@afterenddocumenthook}{}
\def\switch@array{\let\@array@sw\@array@sw@LaTeX}
\makeatother
\usepackage[utf8]{inputenc}
\usepackage[english]{babel}
\usepackage[T1]{fontenc}
\usepackage{amsmath}
\usepackage{hyperref}
\usepackage{microtype}
\usepackage[numbers,sort&compress]{natbib}
\usepackage{graphicx}
\usepackage{tikz}
\usepackage{amsfonts}
\usepackage{braket}
\usepackage{xcolor}
\usepackage{listings}
\usepackage{booktabs}
\usepackage{colortbl}
\usepackage{amssymb}

\definecolor{codebackground}{HTML}{F7F8FA}
\definecolor{codecomment}{HTML}{64748B}
\definecolor{codekeyword}{HTML}{1D4ED8}
\definecolor{coderule}{HTML}{CBD5E1}
\definecolor{ppvmhighlight}{HTML}{EAF1FB}

\lstdefinelanguage{PPVM}{
  alsoletter={._:},
  morekeywords={
    device,fn,
    cpu::cpu.const,cpu::cpu.label,cpu::cpu.br,cpu::cpu.cond_br,cpu::cpu.ret,
    cpu::cpu.add,cpu::cpu.eq,cpu::cpu.lt,cpu::cpu.or,cpu::cpu.dup,
    cpu::cpu.load,cpu::cpu.store,cpu::cpu.print,cpu::cpu.breakpoint,
    circuit::circuit.h,circuit::circuit.x,circuit::circuit.rx,circuit::circuit.cnot,
    circuit::circuit.loss,circuit::circuit.measure,circuit::circuit.reset
  },
  morecomment=[l]{//},
  morestring=[b]",
  sensitive=true
}

\lstdefinelanguage{Python}{
  keywords={False,None,True,and,as,assert,async,await,break,class,continue,
    def,del,elif,else,except,finally,for,from,global,if,import,in,is,
    lambda,nonlocal,not,or,pass,raise,return,try,while,with,yield},
  sensitive=true,
  comment=[l]\#,
  morestring=[b]',
  morestring=[b]",
  morestring=[s]{'''}{'''},
  morestring=[s]{"""}{"""},
}

\newcommand{\fref}[1]{Fig.~\ref{#1}}

\makeatletter
\newcommand\blfootnote[1]{%
  \begingroup
  \renewcommand\thefootnote{}%
  \renewcommand\@makefntext[1]{\parindent\z@\noindent##1}%
  \footnote{#1}%
  \addtocounter{footnote}{-1}%
  \endgroup
}
\makeatother

\begin{document}

\title{PPVM: A Circuit Simulation Framework for Digital Twins of Quantum Processors}

\author{Xiu-Zhe Luo\textsuperscript{*}}
\affiliation{QuEra Computing Inc., 1380 Soldiers Field Road, Boston, MA 02135, USA}
\affiliation{QuEra Computing UK Ltd., Unit 2, Tech Foundry 2, Didcot, OX11 0DF}
\orcid{0000-0002-2445-2701}
\email{rluo@quera.com}

\author{David Plankensteiner\textsuperscript{*}}
\affiliation{QuEra Computing Inc., 1380 Soldiers Field Road, Boston, MA 02135, USA}
\email{dplankensteiner@quera.com}

\author{Kai-Hsin Wu}
\affiliation{QuEra Computing Inc., 1380 Soldiers Field Road, Boston, MA 02135, USA}

\author{Rob Patterson}
\affiliation{QuEra Computing Inc., 1380 Soldiers Field Road, Boston, MA 02135, USA}

\author{Alexander Schuckert}
\affiliation{QuEra Computing Inc., 1380 Soldiers Field Road, Boston, MA 02135, USA}
\affiliation{QuEra Computing UK Ltd., Unit 2, Tech Foundry 2, Didcot, OX11 0DF}
\affiliation{DIENS, École Normale Supérieure, PSL University, CNRS, INRIA, 45 rue d'Ulm, Paris 75005, France}

\author{Oriol Rubies-Bigorda}
\affiliation{QuEra Computing Inc., 1380 Soldiers Field Road, Boston, MA 02135, USA}

\author{Refaat Ismail}
\affiliation{QuEra Computing Inc., 1380 Soldiers Field Road, Boston, MA 02135, USA}

\author{Jonhas Colina}
\affiliation{QuEra Computing Inc., 1380 Soldiers Field Road, Boston, MA 02135, USA}

\author{Neelay Fruitwala}
\affiliation{QuEra Computing Inc., 1380 Soldiers Field Road, Boston, MA 02135, USA}

\author{Danylo Lykov}
\affiliation{NVIDIA Corp, Santa Clara, CA, USA}

\author{Tyson Jones}
\affiliation{NVIDIA Corp, Santa Clara, CA, USA}

\author{Stefan Krastanov}
\affiliation{QuEra Computing Inc., 1380 Soldiers Field Road, Boston, MA 02135, USA}
\affiliation{University of Massachusetts Amherst,
Amherst, USA}

\author{Rafael Haenel}
\affiliation{QuEra Computing Inc., 1380 Soldiers Field Road, Boston, MA 02135, USA}

\author{Stefan Ostermann}
\affiliation{QuEra Computing Inc., 1380 Soldiers Field Road, Boston, MA 02135, USA}

\author{Sheng-Tao Wang}
\affiliation{QuEra Computing Inc., 1380 Soldiers Field Road, Boston, MA 02135, USA}
\orcid{0000-0003-1403-5901}

\maketitle
\blfootnote{\textsuperscript{*}Equal contribution.}

\begin{abstract}
  Quantum computers are approaching the early fault-tolerant regime.
  Verifying the next generation of quantum hardware requires joint emulation of the quantum processor, hardware noise, and classical control and decoding logic.
  We introduce the Programmable Pauli Virtual Machine (PPVM), an open-source framework for digital twins of quantum processors.
  PPVM unifies Pauli propagation and generalized tableau simulation under a common abstraction within a single virtual machine.
  Its bytecode supports arbitrary classical control flow, with measurement-conditioned feed-forward available in the generalized tableau backend.
  For neutral-atom hardware, we derive independent and correlated atom-loss channels and integrate them into both backends.
  We demonstrate loss-aware magic state distillation on over a thousand physical qubits and integration with classical decoding through Bloqade QEC Studio.
\end{abstract}

\section{Introduction}
\label{sec:intro}

Quantum hardware is rapidly approaching the regime of early fault-tolerant quantum computation (FTQC). Verifying architectures designed for early FTQC~\cite{bluvstein2026fault,chiu2025continuous} requires three capabilities: large-scale near-Clifford simulation, faithful modeling of hardware noise, and support for classical control logic. Such logic uses measurement outcomes, detected atom loss~\cite{baranes2026loss}, or decoder output~\cite{ryananderson2021realtime} to determine subsequent operations. Recent work~\cite{li2025soft,haenel2026tsimfastuniversalsimulator,chase2026clifftfastexactsimulation,fang2026symft,aziz2025lowmagic,tuloup2026pfsr,pineiroorioli2026optimizedmps,crognaletti2026mpstab} has made significant progress in near-Clifford simulation using a variety of approaches. Integrating these methods with general classical control and realistic hardware noise in a single framework, however, has received less attention.

\begin{figure*}[!t]
\centering
\includegraphics[width=\textwidth]{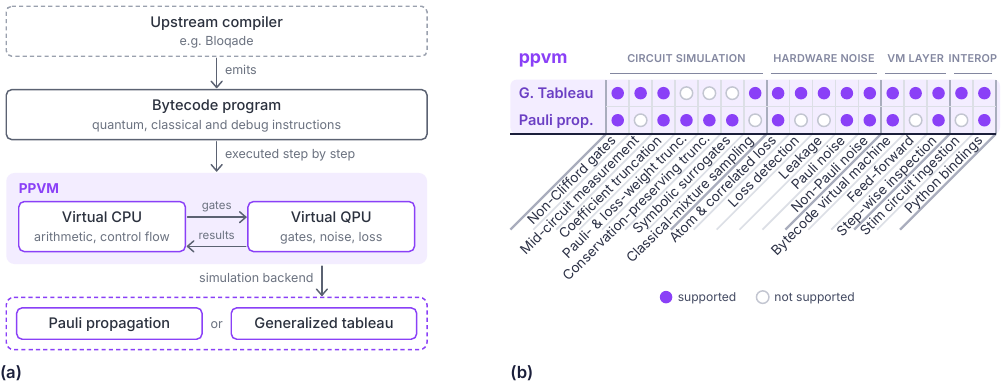}
\caption{\label{fig:overview}
\emph{PPVM overview.} (a) An upstream compiler emits bytecode containing quantum, classical, and debug instructions. The virtual CPU executes arithmetic and control flow and dispatches quantum instructions to the virtual QPU. The virtual QPU applies gates and hardware noise, including independent and correlated atom loss, using either generalized tableau simulation or Pauli propagation. The tableau backend returns measurement outcomes for feed-forward, while the Pauli propagation backend returns expectation values. Breakpoints pause execution for inspection of the classical execution state and the quantum state or propagated observable. (b) Features supported by generalized tableau simulation (G. Tableau) and Pauli propagation (Pauli prop.). Filled and open circles indicate supported and unsupported features, respectively.}
\end{figure*}

For neutral-atom platforms, atom loss represents one of the dominant error channels~\cite{chow2024leakage,bluvstein2026fault}, making accurate loss modeling critical for verifying these architectures. We model both independent loss and correlated loss during two-qubit gates to capture gate-dependent atomic dynamics~\cite{wu2022erasure}. In trajectory-based simulation, a classical bit can track whether each atom has been lost, and subsequent quantum operations can be conditioned on that bit~\cite{suchara2015leakage}. This is one example of the classical state required by realistic quantum programs, which interleave quantum operations with classical logic such as measurement-conditioned gates and loops~\cite{cross2022openqasm,cudaq,ittah2024catalyst,seidel2024qrisp,bloqade_sdk,qir_adaptive}. Faithfully emulating these programs therefore requires tracking quantum evolution alongside the state of the classical controller. Modeling the controller as a virtual machine provides an execution environment with its own instruction set and explicit machine state~\cite{smith2016quil}, allowing programs to run step by step and to be paused and inspected.
PPVM adopts this execution model.
Rather than only simulating a quantum circuit, it emulates the hardware that executes it: a quantum processor as well as its classical control system.

To support adaptive execution, a simulator must retain and update the quantum state as classical control determines subsequent operations. Generalized tableau simulation~\cite{yoder2012generalization,li2025soft} supports this adaptive execution and can be integrated with the classical virtual machine. Although it avoids storing a full state vector, its cost can grow exponentially with the number of non-Clifford gates~\cite{bravyi2016improved}. Pauli propagation~\cite{rudolph2025pauli} offers a complementary approach, evolving observables through the adjoint circuit with cost controlled by truncation. Atom loss can be incorporated through adjoint loss channels, while extending Pauli propagation to measurement-conditioned classical control remains an open problem.

PPVM brings generalized tableau simulation and Pauli propagation together through a backend-agnostic interface that unifies their common components within a shared virtual machine. The two backends differ in their supported features (\fref{fig:overview}b). The framework combines quantum evolution, hardware noise, and classical control in an inspectable execution model, providing a baseline implementation of the simulation requirements for digital twins of quantum processors.

\section{PPVM in a nutshell}
\label{sec:overview}

PPVM emulates a quantum processor and its classical controller as part of a digital twin, with a focus on neutral-atom hardware. It accepts bytecode emitted by an upstream compiler. The bytecode combines classical and quantum instructions with debug instructions for inspecting the full machine state. The workflow is shown in \fref{fig:overview}(a).

For efficient quantum circuit simulation, it unifies two complementary backends in the same virtual machine.
Generalized tableau simulation samples measurement outcomes of large near-Clifford circuits and supports measurement-conditioned feed-forward.
Pauli propagation approximates expectation values of large noisy circuits, even far from the Clifford regime.
Existing simulators generally commit to one of the two paradigms~\cite{li2025soft,haenel2026tsimfastuniversalsimulator,chase2026clifftfastexactsimulation,gidney2021stim,rudolph2025pauli}, whereas PPVM's backend-agnostic interface allows both to share one VM.
\fref{fig:overview}(b) summarizes the features PPVM offers in each mode, including atom loss, arbitrary classical control flow, and a shared bytecode virtual machine.

PPVM also provides Rust and Python interfaces.
Classical arithmetic and control flow can be expressed in either language, while PPVM handles the quantum circuit simulation.
Further usage examples are available in the documentation~\cite{ppvm_website}.

\section{The virtual machine}
\label{sec:vm}

\begin{figure}[!htbp]
\centering
\includegraphics[width=\columnwidth]{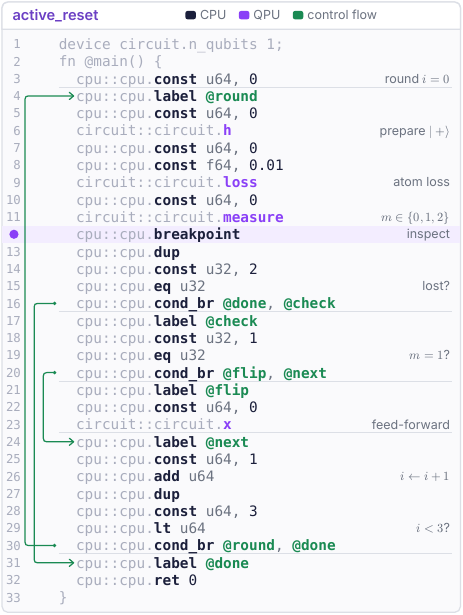}
\caption{\label{fig:control-flow}
\emph{PPVM bytecode program} (text format) for active reset with loss detection.
In each of up to three rounds, the qubit is prepared in $\ket{+}$, subjected to atom loss with probability $1\%$, and measured.
The breakpoint (\textcolor[HTML]{8A40F7}{$\bullet$}) pauses execution immediately after the measurement so that the full machine state can be inspected.
If the atom is lost ($m=2$), the program ends. Otherwise, an $X$ gate is applied when $m=1$, ensuring that a surviving qubit ends the round in $\ket{0}$ (feed-forward).
CPU instructions are shown in black, QPU instructions in purple, and control flow (branch targets and the jumps to them) in green. The loop counter (round) is denoted by $i$.}
\end{figure}

PPVM's virtual machine (VM) executes bytecode, a compact, low-level instruction set that sits between high-level source code and native machine code~\cite{nystrom2021crafting}.
PPVM's bytecode contains classical, quantum, and debug instructions.
The virtual CPU handles integer and floating-point arithmetic and control flow, while quantum instructions, including gates and noise channels, are dispatched to the virtual QPU. The VM uses a last-in, first-out stack as working memory~\cite{nystrom2021crafting}. Values are pushed onto the stack and popped off when consumed. They can also be stored in and loaded from fixed local slots. Gate arguments, such as qubit indices and rotation angles, are popped off the stack, while results (measurement outcomes or expectation values) are pushed onto it.
Because classical and quantum instructions read from and write to the same stack, it bridges classical control flow and quantum operations.
The CPU implements feed-forward by branching on measurement outcomes from the stack. A lost qubit is reported as a distinct measurement outcome, encoded as \texttt{2}, allowing the CPU to branch on loss events. \fref{fig:control-flow} shows an example in the text bytecode format; PPVM also supports a binary representation.

The virtual QPU supports Pauli propagation and generalized tableau simulation, with the tableau backend used by default.
A \texttt{device} directive in the bytecode header selects the backend for each program.
The tableau backend evolves the quantum state in the Schr\"odinger picture.
Measurement outcomes are sampled and pushed onto the stack, enabling feed-forward as illustrated in \fref{fig:control-flow}.

The Pauli propagation backend instead evolves an observable $\mathcal{O}$, specified in the bytecode header, through the adjoint circuit in the Heisenberg picture. A \texttt{trace} instruction computes the overlap of the evolved observable with the initial state and pushes the resulting expectation value $\braket{\mathcal{O}}$ onto the stack. This avoids bitstring sampling. This mode does not support sampled mid-circuit measurements, measurement-based resets, or measurement-conditioned feed-forward. Classical loops and parameters can specify the circuit to be propagated. Incorporating arbitrary classical control into backward operator evolution remains an open problem, although some feed-forward operations can be expressed as channels via their Kraus decomposition.

The virtual CPU and QPU are implemented as standalone components using the \texttt{vihaco} (\textbf{Vi}rtual \textbf{Ha}rdware \textbf{Co}mponents) framework~\cite{vihaco}. Each component declares its own instruction set, from which the text parser and binary encoding are derived. \texttt{vihaco} composes these components into a single virtual machine, allowing additional virtual components, such as a real-time decoder, to be added without modifying the existing components. Classical control flow can depend on decoded logical outcomes as well as direct measurement results. Bloqade QEC Studio demonstrates this feedback through integration with the \texttt{deq} decoding framework~\cite{Microsoft_QDK_EC} and provides interactive examples of adaptive quantum programs~\cite{qec-studio}.

Step-wise execution also supports standard debugging tools. PPVM provides a debugger for stepping through bytecode instructions and a command-line interface (CLI).
These tools make PPVM a useful compilation target for testing and debugging programs intended for quantum hardware.

\section{Theoretical background of simulation methods}
\label{sec:background}

\subsection{Operator propagation as a shared abstraction}
\label{sec:background:abstraction}

Near-Clifford simulations often describe features of the quantum system in terms of Pauli strings (tensor products of Pauli operators).
The $n$-qubit Pauli group is defined as
\begin{align}
\label{eq:pauli_group}
    \mathcal{P}_n = \left\{i^k P_1P_2 ... P_n \right\},
\end{align}
where $k\in\{0,1,2,3\}$ and $P_j\in\{I_j, X_j, Y_j, Z_j\}$ is one of the Pauli operators acting on the $j$th qubit.
We omit tensor-product symbols, which are implicit in the qubit indices. We call the $4^n$ elements with $k=0$ \textit{Pauli words}, and general elements $Q\in\mathcal{P}_n$ \textit{phased Pauli words}.

Efficient simulation in this formalism rests on a key property of Clifford operations: conjugation maps each Pauli-group element to another Pauli-group element \cite{gottesman1998heisenberg}.
Let $C$ be a Clifford operation applied to a phased Pauli word $Q$.
Then
\begin{align}
\label{eq:clifford_conjugation}
    C Q C^\dagger = Q^\prime,
\end{align}
where $Q^\prime\in\mathcal{P}_n$.

\subsection{Pauli propagation}

In Pauli propagation approaches~\cite{rudolph2025pauli}, rather than tracking the explicit state of a system, we track an observable of interest $\mathcal{O}$, for which we seek the expectation value.
Since the $4^n$ Pauli words $P_1P_2\dots P_n$ form a basis of the space of operators acting on the $n$-qubit Hilbert space, any operator can be expressed as a linear combination of phased Pauli words,
\begin{align}
    \label{eq:pauli_sum}
    \mathcal{O} = \sum_i c_i Q_i,
\end{align}
where $c_i\in\mathbb{R}$ and $Q_i\in \mathcal{P}_n$.

Let $\mathcal{E}$ be a completely positive trace-preserving (CPTP) map that describes a noisy quantum circuit.
The final (mixed) state $\rho$ of the system at the end of the circuit is
\begin{align}
    \rho = \mathcal{E}[\rho_0].
\end{align}
Unless otherwise stated, we take $\rho_0 = \left(\ket{0}\bra{0}\right)^{\otimes n}$.

The expectation value of the observable $\mathcal{O}$ after the circuit is given by
\begin{align}
    \braket{\mathcal{O}} = \mathrm{tr}\left(\mathcal{O}\rho\right) = \mathrm{tr}\left(\mathcal{O}\mathcal{E}[\rho_0]\right) = \mathrm{tr}\left(\mathcal{E}^\dagger [\mathcal{O}]\rho_0\right),
\end{align}
where in the final step we used the cyclic property of the trace and $\mathcal{E}^\dagger$ is the adjoint circuit.
Therefore, instead of propagating the state through the system, we may propagate an observable backwards through the circuit.

For example, if the circuit is purely unitary, then
\begin{align}
    \mathcal{E}[\rho] = U \rho U^\dagger
        \;\longleftrightarrow\;
    \mathcal{E}^\dagger[\mathcal{O}] = U^\dagger \mathcal{O} U.
\end{align}

At this point, the relation to \eqref{eq:clifford_conjugation} also becomes apparent: if the circuit is purely Clifford, then each operation in the circuit simply conjugates every term in the sum describing $\mathcal{O}$ with the \textit{adjoint} of the operation. By tracking the conjugations, we can efficiently compute $\braket{\mathcal{O}}$ for Clifford circuits.

Non-Clifford operations are treated by decomposing them into sums of Clifford operations and injecting new terms into the sum in \eqref{eq:pauli_sum}.
For example, consider the single-qubit rotation about a Pauli axis by an arbitrary angle. Throughout this work, we use the \emph{half-angle} convention for rotations,
\begin{align}
    \label{eq:pauli_rotation}
    R_P(\theta) = e^{-i\theta P/2} = I \cos{\frac{\theta}{2}} - i P\sin{\frac{\theta}{2}},
\end{align}
where $P\in\{X, Y, Z\}$.

Propagating an observable backwards through such a gate conjugates each phased Pauli word $Q$ as $R_P^\dagger Q R_P$. If $Q$ commutes with $P$, the word is unchanged. If it anticommutes, the result is a sum of two phased Pauli words,
\begin{align}
    \label{eq:pauli_rotation_branch}
    R_P^\dagger\, Q\, R_P = \cos{\theta}\,Q + \sin{\theta}\,\left(iPQ\right),
\end{align}
the original word retaining $\cos{\theta}$ and the new word $iPQ\in\mathcal{P}_n$ carrying $\sin{\theta}$. The new term can be thought of as a new branch in a tree graph. Note that forward state propagation, $R_P Q R_P^\dagger$, carries the opposite sign on the branch term.

While this suffices to describe non-Clifford operations and thereby arbitrary circuits, it has a clear downside: the number of terms in \eqref{eq:pauli_sum} can, in the worst case, grow exponentially with the number of non-Clifford operations. However, the theoretical exponential scaling can often be overcome in practice.
This may be achieved by employing different truncation strategies, which prune terms from the sum under specific conditions.
One such strategy is to neglect terms with coefficient magnitudes below a certain threshold, since these terms are expected to contribute small summands to the final expectation value.
Another strategy is to neglect terms with Hamming weight (the number of non-identity Pauli constituents in the word) exceeding a certain threshold.
Local depolarizing noise exponentially suppresses high-weight Pauli terms, motivating this latter truncation strategy~\cite{schuster2025noisy}.

Since every branch after a rotation only multiplies coefficients by $\cos\theta$ or $\sin\theta$, the same propagation also works for symbolic angles.
Replacing the numerical coefficients $c_i$ by a sine-cosine power-series data structure yields $\braket{\mathcal{O}}$ as a symbolic function of the circuit parameters, useful as a \textit{surrogate}~\cite{rudolph2023classical}.
The symbolic analogue of coefficient truncation is to neglect terms above a chosen order in $\sin\theta$.
For small angles, $\sin\theta \approx \theta$, so this corresponds to a controlled expansion in the rotation angles.

\subsection{Generalized tableau simulation}

Another well-established technique for the efficient simulation of Clifford circuits is the stabilizer tableau~\cite{aaronson2004improved}.
The fundamental idea is that there exists a set of states, the so-called stabilizer states, which are stabilized by a subgroup $S \subset \mathcal{P}_n$; i.e., the states are eigenstates of all $2^n$ elements of $S$ with the eigenvalue +1.

The $n$ generators $s_i$ of $S = \langle s_1, s_2, \dots, s_n\rangle$ can be thought of as phased Pauli words.
In addition to the stabilizers, we also track a set of destabilizers, $\{ d_1, d_2, \dots, d_n\}$, chosen such that $d_j$ anticommutes with $s_j$ and commutes with all other $s_k$ and all $d_k$~\cite{aaronson2004improved}.
To simulate a Clifford gate, we conjugate each stabilizer and destabilizer generator by that gate.

Generalized tableau simulation refers to the extension of tableau simulation beyond Clifford circuits~\cite{yoder2012generalization, li2025soft}.
We summarize the basics here and refer to Ref.~\cite{yoder2012generalization} for a detailed treatment.
The key insight is that applying all possible combinations of the destabilizers to a stabilizer state $\ket{\psi_S}$ generates a basis $\{\ket{b_{\boldsymbol{\alpha}}}\}$ of the full $n$-qubit Hilbert space,
\begin{align}
    \ket{b_{\boldsymbol{\alpha}}} = d_{\boldsymbol{\alpha}} \ket{\psi_S} = \prod_{k=1}^{n} d_k^{\alpha_k} \ket{\psi_S},
\end{align}
where $\alpha_k \in \{0, 1\}$, i.e., $\boldsymbol{\alpha}$ is a length-$n$ bit string.
We can write an arbitrary state in this basis
\begin{align}
    \ket{\phi} = \sum_{\boldsymbol{\alpha}} v_{\boldsymbol{\alpha}} d_{\boldsymbol{\alpha}}\ket{\psi_S},
\end{align}
where $v_{\boldsymbol{\alpha}}\in\mathbb{C}$ are the coefficients of the linear combination.

The coefficient vector $\boldsymbol{v}$ can remain sparse when relatively few non-Clifford gates have been applied to an initial stabilizer state.
In addition to this, we instantiate $\ket{\psi_S}$ as a destabilizer-stabilizer tableau, of size $O(n^2)$.
Applying Clifford gates to $\ket{\phi}$ effects a basis change, which does not affect the coefficient vector~\cite{yoder2012generalization}, but simply updates the tableau according to the Aaronson-Gottesman algorithm~\cite{aaronson2004improved}, which means we conjugate the tableau rows according to \eqref{eq:clifford_conjugation}.
In contrast, when applying a non-Clifford Pauli rotation, the number of non-zero coefficients in $\boldsymbol{v}$ may at most double~\cite{li2025soft}.

Measurements do not grow the number of non-zero coefficients, but may reduce it.
Measuring qubit $j$ in the computational basis distinguishes two cases~\cite{yoder2012generalization}.
If $Z_j$ commutes with all stabilizers, each basis state $\ket{b_{\boldsymbol{\alpha}}}$ is an eigenstate of $Z_j$ with eigenvalue $\lambda_{\boldsymbol{\alpha}} = \pm 1$, so outcome $m \in \{0, 1\}$ occurs with probability
\begin{align}
    p_m = \sum_{\boldsymbol{\alpha}\,:\,\lambda_{\boldsymbol{\alpha}} = (-1)^m} |v_{\boldsymbol{\alpha}}|^2.
\end{align}
The projection discards all other coefficients and renormalizes, leaving the tableau unchanged while possibly reducing the number of non-zero coefficients.
Otherwise, $Z_j$ maps basis states onto each other in pairs, so the projection merges paired coefficients and the tableau is updated following the Aaronson-Gottesman algorithm.

Since Clifford gates and measurements never increase the number of non-zero coefficients, the exponential cost of simulation is shifted to the number of non-Clifford operations in a circuit.

\section{Algorithmic contributions}
\label{sec:algorithms}

We now describe how the simulation backends incorporate atom loss, while maintaining the capabilities required for a digital twin.
We also show methods for conserving observables under truncation and sampling from classical mixtures.

\subsection{Quantum channels for atom loss and reset}

We model atom loss by introducing an additional \emph{loss} state, $\ket{L}$, so that each qubit is described in the basis $\{\ket{0},\ket{1},\ket{L}\}$.
To obtain the channel associated with a given quantum operation, we simulate the full dynamics, including the loss state and all relevant loss processes, over the operation time $t$. By factoring out the ideal unitary evolution, we extract the Kraus operators that define the effective loss channel. The action of a channel $\mathcal{E}$ with Kraus operators $\{K_i \}$ on a density operator $\rho$ is given by
\begin{align}
    \mathcal{E}[\rho] = \sum_i K_i \rho K_i ^\dagger.
\end{align}
When propagating observables the relevant map is the adjoint channel. For an operator $\mathcal{O}$, the corresponding evolution is
\begin{align}
\label{eq:adjoint_kraus}
    \mathcal{E}^\dagger[\mathcal{O}] = \sum_i K_i^\dagger \mathcal{O} K_i.
\end{align}
Loss channels are generally not self-adjoint.
Forward state propagation makes use of $\mathcal{E}$, while backward observable propagation uses $\mathcal{E}^{\dagger}$.

In what follows, we present the Kraus operators describing independent and correlated atom loss processes, as well as the reset channel that returns lost atoms to the computational subspace, $\{\ket{0},\ket{1}\}$. In our model, transitions from the computational subspace to the loss state $\ket{L}$ occur exclusively through incoherent processes with rates $\gamma_0$ and $\gamma_1$. Population in $\ket{L}$ remains decoupled from subsequent dynamics and can only return to the computational subspace through the reset process $\ket{L}\rightarrow\ket{0}$. Consequently, neither loss nor reset generates coherences between the computational and loss subspaces.

\subsubsection{Independent loss}

When loss occurs independently for each atom, the corresponding single-qubit loss channel is described by the Kraus operators
\begin{subequations}
\begin{align}
    K_0 =& \sqrt{1 - p_0}\ket{0}\bra{0} + \notag \\
    & \sqrt{1 - p_1}\ket{1}\bra{1} + \ket{L}\bra{L},\\
    K_1 =& \sqrt{p_0} \ket{L}\bra{0},\\
    K_2 =& \sqrt{p_1} \ket{L}\bra{1},
\end{align}
\end{subequations}
where $p_0$ and $p_1$ are the probabilities of losing a qubit from the computational states $\ket{0}$ and $\ket{1}$, respectively.
For a qubit initially in the computational subspace and in the absence of additional dynamics,
\begin{align}
    p_i = 1 - e ^ {-\gamma_i t}.
\end{align}
In such cases, the total loss probability during the process is
\begin{align}
    p_\mathrm{tot} = p_0\braket{\Pi_0}_\mathrm{init} + p_1\braket{\Pi_1}_\mathrm{init} = \braket{L}_\mathrm{final},
\end{align}
where we have defined the projectors $\Pi_i = \ket{i}\bra{i}$ and $L = \ket{L}\bra{L}$. Here, $\braket{\cdot}_\mathrm{init}$ and $\braket{\cdot}_\mathrm{final}$ denote expectation values before and after application of the channel, respectively.

This channel allows for state-dependent loss, such that $p_0$ and $p_1$ may differ. Such asymmetries can occur in neutral-atom gate implementations~\cite{evered2023gates,sahay2023biased} and can be important for accurately modeling loss processes and studying effects such as post-selection bias. In many applications, however, it is sufficient to consider the symmetric approximation, $p_0 = p_1 = p_\mathrm{tot}$.

\subsubsection{Correlated loss}
\label{sec:correlated-loss}

When two qubits undergo joint dynamics, as is the case during a two-qubit gate, loss processes may become correlated~\cite{wu2022erasure}. We distinguish between two situations.

In the first case, both qubits are initially in the computational subspace. During the gate, either both qubits are lost with probability $p_{LL}$, or each qubit is lost individually with probability $p_{LQ}$. Thus, the probability that both qubits remain in the computational subspace after the gate is $1-p_{LL}-2p_{LQ}$.

In the second case, one qubit has already been lost prior to the gate. The remaining qubit then undergoes dynamics that generally differ from those of a gate acting on two present atoms~\cite{bluvstein2026fault}. As a result, its probability of being lost during the gate, $p_{LN}$, may also differ.

For more details on the Kraus representation of the correlated loss channel, see Appendix~\ref{app:correlated-loss}.

\subsubsection{Loss reset channel}

To simulate hardware with loss detection capabilities, we also model the reset of a lost qubit into the computational subspace.
We describe this process by the reset channel
\begin{subequations}
\begin{align}
    K_0 &= \ket{0}\bra{0} + \ket{1}\bra{1},\\
    K_1 &= \ket{0}\bra{L},
\end{align}
\end{subequations}
which returns all population in the loss state to the $\ket{0}$ state while leaving the computational subspace unchanged.

\subsection{Atom loss in Pauli propagation}

In order to describe atom loss in Pauli propagation, we extend the Pauli basis.
A general description of a three-level system would require a basis of nine orthonormal operators.
Since neither the quantum operations nor the noise channels considered here generate coherences between the computational subspace and the loss state, the dynamics can instead be faithfully represented by augmenting the Pauli basis with the projector on the loss state, $L$, only.
The resulting single-qubit Pauli basis including loss is then given by the set of $3\times 3$ matrices, $\{I, X, Y, Z, L\}$.
To maintain orthogonality, the $2 \times 2$ Pauli operators are embedded into the enlarged Hilbert space by extending them with a row and column of zeros, e.g., $I=\text{diag}(1, 1,0)$, $Z = \text{diag}(1, -1, 0)$ and $L = \text{diag}(0,0,1)$. A subtle but important distinction arises when representing gates in the extended basis. Since a lost qubit is unaffected by subsequent operations, gates must leave the loss projector invariant. Consequently, the matrix representation of a gate differs from the corresponding basis element. For example, the $Z$ gate acts as $Z_\mathrm{gate} = \text{diag}(1, -1, 1) \neq Z$.
The unit entry in the loss subspace reflects the fact that the gate acts trivially on a lost qubit.

\subsubsection{Independent loss}

Using the Kraus operators for the independent loss channel together with the adjoint mapping in~\eqref{eq:adjoint_kraus}, we obtain the action of the loss channel on the extended Pauli basis.
In the symmetric case, $p_0=p_1=p_\mathrm{tot}\equiv p$, the adjoint channel acts as
\begin{subequations}
\begin{align}
    \label{eq:pp_loss_action_1}
    \mathcal{E}^\dagger[P] &= (1 - p) P, ~ P\in\{I, X, Y, Z\},\\
    \label{eq:pp_loss_action_2}
    \mathcal{E}^\dagger[L] &= L + p I.
\end{align}
\end{subequations}
Equation \eqref{eq:pp_loss_action_1} shows that atom loss uniformly suppresses observables in the computational subspace, while \eqref{eq:pp_loss_action_2} accounts for the corresponding accumulation of population in the loss state.

Note that if we start with an observable containing no $L$ factors, there is no $L$ character present in any Pauli word forming the observable.
Since the loss channel does not map Pauli operators in the computational subspace onto the loss projector $L$, Pauli propagation remains entirely in the computational subspace and extending the Pauli basis becomes unnecessary.
However, the extension is needed in the presence of reset operations (which map the loss state back to the computational subspace) or if a measurement falsely detects a lost atom as being in $\ket{0}$ or $\ket{1}$ -- these cases can also be captured by applying a reset-to-$\ket{0}$ or reset-to-$\ket{1}$ channel, respectively, immediately before measurement.

While the extension to asymmetric loss ($p_0\neq p_1$) is straightforward, its practical use is limited, as the adjoint channel no longer acts as a pure rescaling on the computational subspace.
For example, the identity branches as
\begin{align}
    \mathcal{E}^\dagger[I] = \left(1 - \frac{p_0 + p_1}{2}\right) I + \frac{p_1 - p_0}{2} Z.
\end{align}
Since the branching applies to identity factors, loss channels can generate branches even outside the support of the observable and without any reset operations.
Therefore, asymmetric loss can quickly lead to a blow-up in the number of branches.

\subsubsection{Correlated loss}

The action of the correlated loss channel takes the more complicated form
\begin{subequations}
    \begin{align}
        \label{eq:pp_correlated_loss_action_1}
        \mathcal{E}^\dagger[P_1P_2] &= (1 - 2p_{LQ} - p_{LL})P_1P_2,\\
        \mathcal{E}^\dagger[PL] &= p_{LQ}PI + (1 - p_{LN})PL,\\
        \mathcal{E}^\dagger[LL] &= LL + p_{LN}(IL + LI) + p_{LL} II,
    \end{align}
\end{subequations}
where $P, P_1, P_2 \in \{I, X, Y, Z\}$.
This representation enables the simulation of two-qubit correlated loss channels that more closely resemble hardware-level noise, at the cost of an increased number of branches generated when propagating observables.

\subsubsection{Loss reset}

The adjoint action of the reset channel is
\begin{subequations}
\begin{align}
    \mathcal{E}^\dagger[P] &= P, ~ P\in\{X, Y\},\\
    \label{eq:pp_loss_reset_2}
    \mathcal{E}^\dagger[P] &= P + L, ~ P\in \{I, Z\},\\
    \mathcal{E}^\dagger[L] &= 0.
\end{align}
\end{subequations}
At a glance, it may seem counterintuitive that the reset channel generates terms proportional to the loss projector $L$ via \eqref{eq:pp_loss_reset_2}.
This is a consequence of working in the Heisenberg picture: under backward propagation, the reset channel is applied \textit{before} the preceding loss channels.
It is the combination of the reset channel \eqref{eq:pp_loss_reset_2} followed by the loss channel \eqref{eq:pp_loss_action_2} that reproduces the effect of resetting lost population to the computational subspace.
If no loss channel precedes the reset operation, every Pauli word containing an $L$ character introduced by the reset channel retains that $L$ throughout the subsequent backwards propagation.
Since the initial state lies entirely within the computational subspace, these Pauli words have zero overlap with the initial state and therefore do not contribute to any expectation values.

\subsubsection{Truncating high-loss branches}

Several of the noise channels introduced above lead to branching and thus, in principle, to an exponential growth in the number of terms in an observable $\mathcal{O}$.
In practice, however, this exponential scaling rarely poses a problem.
First off, note that for symmetric loss and observables initially containing no $L$ factors, branching on loss channels occurs only when reset operations are present.
Without reset channels, no $L$ character is introduced in any Pauli word and neither \eqref{eq:pp_loss_action_1} nor \eqref{eq:pp_correlated_loss_action_1} generates branching.

Next, let us assume that there is a significant number of reset channels applied throughout the circuit.
This will lead to a number of words with multiple $L$ characters.
However, these terms do not contribute directly to the final expectation value, since their overlap with any initial state in the qubit subspace vanishes.
For independent symmetric loss in a single layer, replacing all $k$ loss projectors by computational-subspace identities introduces a factor $p^k$. With correlated loss, however, a single channel can remove two loss projectors with a factor $p_{LL}$ through the $LL\rightarrow II$ term. While the loss probabilities are usually small in practice, that alone does not guarantee negligible aggregate contributions.

We therefore use a configurable cutoff on the number of loss projectors to control the growth of loss-related branches. This truncation does not by itself guarantee polynomial complexity, since non-Clifford operations can still generate exponentially many terms.

\subsection{Atom loss in generalized tableaux}
In generalized tableaux, loss can be tracked with additional classical flags, in lieu of extending the basis.
Recall that there are no coherences between the computational subspace $\{\ket{0},\ket{1}\}$ and the loss state $\ket{L}$.
Therefore, in a single shot, each qubit is either entirely in the loss state $\ket{L}$ if it has been lost, or entirely in the computational subspace if no loss event occurred.
The occupation of $\ket{L}$ can hence be represented by a Boolean variable, which is zero initially, but flips to one if a qubit experiences a loss event.
A reset channel acting on a lost qubit simply flips the Boolean value back to zero and resets the qubit to the $\ket{0}$ state.
Averaging the Boolean value over many shots estimates the expectation value $\braket{L}$, which can also be computed using Pauli propagation.

This representation of loss is both simple and computationally cheap: it requires only a list of $n$ Boolean values, one for each qubit, that stores the information about whether a qubit has been lost.
It also naturally supports the simulation of architectures capable of loss detection, since the loss information is readily available.
Each executed gate or noise channel checks the loss flags and uses the appropriate dynamics for that loss configuration. Reset operations return lost qubits to the computational subspace.
For the loss channels implemented here, loss outcomes are sampled according to the channel probabilities, and the quantum state and classical loss flags are updated for each outcome.

\subsection{Leakage}

Leakage is a process similar to loss.
It describes the process of a qubit being driven or decaying out of the computational subspace.
In our model, leaked qubits remain indistinguishable from one of the computational states in a measurement and are decoupled from subsequent gates.
This captures, for instance, leakage of neutral atoms into other ground-state hyperfine levels, which do not couple to the Rydberg laser and are not revealed by loss detection~\cite{bluvstein2026fault}.
Their measurements return ordinary binary outcomes.
At the time of writing, leakage is only supported in generalized tableau simulations.
Its representation is similar to that of loss.
We track the state of a qubit (active, lost, or leaked) using a classical, ternary flag.
Both leaked and lost atoms skip gates, but leaked atoms are measured as if they were in the computational subspace.

\subsection{Enforcing observable conservation}
\label{sec:conservation}

Symmetry and conserved quantities are often used in simulation techniques to ensure correctness or reduce computational complexity. One class of such problems involves Hamiltonians, such as the Heisenberg and XY models, which conserve the total magnetization $M=\sum_i Z_i$.
In PPVM, we make it possible to preserve overlap with conserved quantities by augmenting the truncation with a \emph{protected set}: a user-specified collection of Pauli words that are exempt from truncation. This overlap is preserved exactly if the protected set contains all Pauli words in the conserved quantity's expansion and truncation is performed only after complete gate blocks that conserve that quantity. For the total magnetization $M$, protecting the single-site Pauli words $\{Z_i\}$ suffices.
The computational cost is usually low, as we only need to store and consult a tractable list of protected operators.

\subsection{Efficient, approximate sampling with classical mixtures}\label{sec:classical-mixture}

We can represent specific types of mixed states as a sum over pure state projectors, where each pure state is represented by a generalized tableau.
As in Pauli propagation, this representation supports branching and truncation, though the number of induced branches differs between the stabilizer and Pauli bases.
A mixed state $\rho$, that is a classical mixture of pure states, can be written as
\begin{align}
    \label{eq:classical_mixture}
    \rho = \sum_k p_k \ket{\phi_k}\bra{\phi_k},
\end{align}
where the system is found in the pure state $\ket{\phi_k}$ with probability $p_k$.

If the initial state is pure, then we start with only one term in the sum in \eqref{eq:classical_mixture}.
A probabilistic channel $\mathcal{E}$ that applies a noise process $\mathcal{D}$ with a probability $p$ inserts new terms in the sum,
\begin{align}\label{eq:classical_mixture_branching}
    \mathcal{E}[\rho] = (1 - p) \rho + p \mathcal{D}[\rho].
\end{align}
For example, a simple bit-flip error $\mathcal{E}_X$ that occurs with the probability $p_X$ results in one additional summand,
\begin{align}
    \mathcal{E}_X[\rho] = (1 - p_X)\rho + p_X X \rho X,
\end{align}
whereas a depolarization error inserts three new summands,
\begin{align}
    \mathcal{E}_\mathrm{dpl} [\rho] = (1 - p)\rho + \frac{p}{3}\left(X\rho X + Y\rho Y + Z\rho Z\right).
\end{align}
Such channels induce branching in generalized tableaux, though they are diagonal in Pauli propagation.
Measurements also create classical mixtures, mixing the two possible outcomes.
Consider the quantum state averaged over the outcomes of a binary projective measurement,
\begin{align}
    \mathcal{M}[\rho] = \Pi_0\rho \Pi_0 + \Pi_1\rho \Pi_1,
\end{align}
where $\Pi_i$ are the outcome projectors, with outcome probabilities $\mathrm{tr}(\Pi_i\rho)$.
We can, in principle, account for feed-forward by retaining the measurement outcome in each branch's classical state and applying the corresponding conditional gate.

If we choose to represent the pure states by generalized tableaux and their coefficient vectors remain sparse, applying projectors and gates is efficient.
With this, we can propagate a state containing probabilistic operations such as mid-circuit measurements and noise channels.
The final mixture represents an average over circuit realizations, subject to any truncation.
Sampling bit strings from the final state is then efficient.
We choose one of the pure states in the classical mixture randomly, weighted by the respective probabilities, and then apply a measurement to sample a bit string according to the generalized tableau algorithm~\cite{yoder2012generalization, li2025soft}.
The resulting data structure is similar to the concept of \texttt{petrajectories} in QuantumClifford.jl~\cite{krastanov2026quantumclifford}.

Each probabilistic operation can create new branches in the classical mixture from \eqref{eq:classical_mixture}.
Before merging, each child branch's weight equals its parent's weight multiplied by a conditional outcome probability, so it cannot exceed the parent's weight.
Unlike Pauli expansion coefficients, the mixture weights are nonnegative probabilities and must satisfy
\begin{align}
    \sum_k p_k = 1.
\end{align}
Although the number of branches can grow exponentially with the number of probabilistic operations, low-probability branches can be truncated.
Choosing some small cutoff $0<C\ll1$, we keep only terms where $p_k>C$ and renormalize the retained weights, provided at least one term remains.
This limits the number of terms in \eqref{eq:classical_mixture} to at most $1/C$.
Accuracy depends on the total probability discarded over the evolution.

As a simplified example, say we apply $K$ independent bit-flip channels, each with the same probability $p < 1/2$.
Without merging equivalent branches, the overall number of branches is $2^K$.
If the cutoff is chosen to retain exactly the branches with at most $r$ bit-flip events, the number of retained branches is
\begin{align}
    \label{eq:binomial_scaling}
    \sum_{j = 0} ^ r \binom{K}{j}.
\end{align}

The overall scaling of this approach is still exponential in the number of non-Clifford gates as each pure state in the classical mixture is represented by a generalized tableau.
However, depending on the noise probabilities and the chosen cutoff, we can avoid paying the full exponential scaling in the number of probabilistic operations.
Still, the scaling shown in \eqref{eq:binomial_scaling} can quickly lead to an intractable number of branches.
We note, however, that this approach allows for efficient sampling in a noisy variant of the $[[17,1,5]]$ color-code (85 qubits) magic state distillation circuit~\cite{sales2025experimental}.

\section{Demonstrations}
\label{sec:demos}

\subsection{Loss-aware magic state distillation}

We simulate a magic state distillation (MSD) scheme from Ref.~\cite{sales2025experimental} including atom loss alongside depolarization noise.
We perform a 5-to-1 distillation of the Bravyi-Kitaev $T$-state~\cite{Bravyi2005}.
To study the scalability, we encode the distillation circuit in the family of triangular 4.8.8 color codes~\cite{landahl2011fault}, $[[(d^2-1)/2+d,\,1,\,d]]$ for odd code distance $d$.
Its smallest members are the $[[7,1,3]]$ Steane code and the $[[17,1,5]]$ color code used
in Ref.~\cite{sales2025experimental}.
We scale it up to $d=23$, i.e.\ $[[287,1,23]]$.
The stabilizers are generated with the Performance Estimator of Codes On Surfaces (PECOS)~\cite{pecos}, and each magic state is injected
with a unitary encoding circuit obtained from Stim's generic stabilizer-tableau
synthesis~\cite{gidney2021stim}, without further optimization.
This gives a total of $5n$ physical qubits, from $35$ to $1435$, where $n$ is the number of physical qubits per code block.

\begin{figure}[!htbp]
    \centering
    \includegraphics[width=\columnwidth]{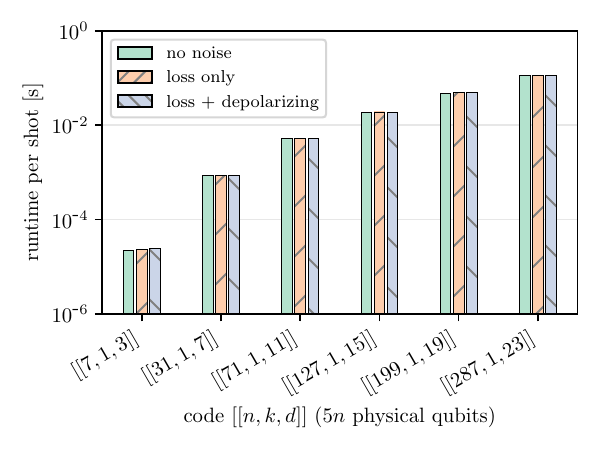}
    \caption{\label{fig:msd-runtime}
        \emph{Runtime of loss-aware MSD with different codes.} For the code sizes and noise rates shown, adding loss and depolarizing noise produces only a small runtime overhead. At a total of $1435$ qubits, the runtime per shot is about $0.1\mathrm{s}$. The noise probabilities are $p_{\mathrm{loss}} = p_{\mathrm{dpl}} = 10^{-4}$. All runtimes were measured via PPVM's Python bindings on a single thread of an Apple M4 (24 GB RAM).}
\end{figure}

The noise model is intentionally kept simple for this demonstration.
Each single-qubit gate is followed by a single-qubit depolarization channel and an independent loss channel, with probabilities $p_{\mathrm{dpl}}$ and $p_{\mathrm{loss}}$, respectively.
Every two-qubit gate is followed by a two-qubit depolarization channel (again with $p_{\mathrm{dpl}}$) and a correlated loss channel with $p_{LL} = p_{LN} = p_{LQ} = p_{\mathrm{loss}} / 3$.
Simulation runtimes are shown in Figure~\ref{fig:msd-runtime}.

\subsection{Noisy QEC simulation with real-time decoding}

PPVM serves as the simulation backend of Bloqade QEC Studio~\cite{qec-studio}, a system-level environment for simulating adaptive logical quantum programs.
It integrates PPVM and the \texttt{deq} decoding framework~\cite{Microsoft_QDK_EC}.
Users assemble a QEC experiment as a control flow graph of gadgets (state preparation, syndrome-extraction rounds, logical gates, and readout).
The underlying circuits are simulated using PPVM and can therefore include non-Clifford gates and qubit loss.
The integration with \texttt{deq} is made possible by the execution model described in Sec.~\ref{sec:vm}: the simulator state persists across gadgets and measurement results are available for decoding before execution continues.
The classical control flow can therefore branch on decoded logical outcomes. Execution then continues from the quantum state retained across gadgets, closing the decoding and control feedback loop.
The current integration simulates atom loss without using loss reports in decoding.

Bloqade QEC Studio provides a number of example programs, including magic state distillation with the $[[17,1,5]]$ color code~\cite{sales2025experimental} and resource state preparation using the space-time efficient analog rotation (STAR) protocol on self-dual bivariate bicycle (BB) codes~\cite{ismail2026fastparallelhighratestar}.

\subsection{Effects of loss on Heisenberg dynamics}
\label{sec:transport}

We demonstrate conservation under truncation using protected Pauli strings in a many-body system subject to loss. The spin-$1/2$ Heisenberg model is defined by the Hamiltonian
\begin{equation}
    \label{eq:heisenberg}
  H = \frac{J}{2}\sum_{\langle i,j\rangle}\big(X_iX_j + Y_iY_j + Z_iZ_j\big)
\end{equation}
on an $\ell\times \ell$ square lattice, which conserves the total magnetization
$M=\sum_q Z_q$. We seed the Heisenberg-picture observable with a single $Z_0$ on
the central site, propagate it backwards through a
first-order Trotter circuit, and evaluate its overlap with the total magnetization,
\begin{align}
  \label{eq:transport_obs}
  m(t) &= \frac{1}{2^n}\mathrm{tr}\left[MZ_0(t)\right] \\
       &= \sum_q \frac{1}{2^n}\mathrm{tr}\left[Z_qZ_0(t)\right], \notag
\end{align}
where $n=\ell^2$ is the number of qubits. By Pauli orthogonality, $m(t)$ is the sum of the coefficients of the single-site $Z_q$ terms in the propagated observable. Each term in the sum is an
infinite-temperature spin correlation function, as commonly studied in the
context of spin transport~\cite{rakovszky2022daoe, PhysRevB.101.020416}. Here, each $Z_q$ is embedded with the computational-subspace identity $I=\mathrm{diag}(1,1,0)$ on all other sites. These operators vanish whenever any atom is lost. Thus, in the presence of loss, $m(t)$ is a spin-correlation sum weighted by survival of the entire register.

\begin{figure}[!t]
    \centering
    \includegraphics[width=\columnwidth]{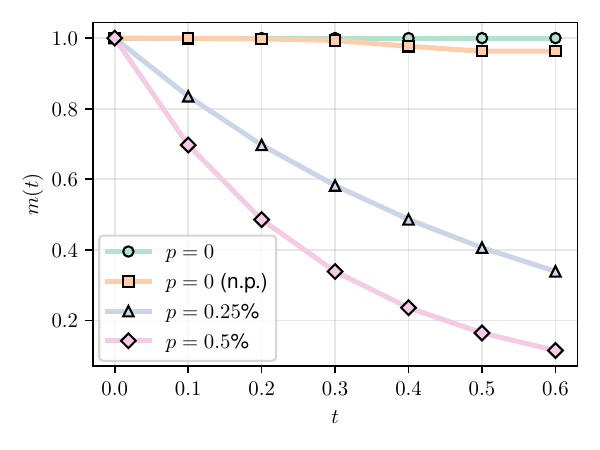}
    \caption{\label{fig:transport}
    \emph{Loss-induced decay of the summed spin correlation $m(t)$ in the Heisenberg model.} $\ell=3$ square
    lattice ($n=9$ qubits), $J=1$, $dt=0.1$, $6$ first-order Trotter steps. Each
    two-qubit interaction is decomposed into single-qubit Cliffords, $R_Z$ rotations, and CZ
    gates, with a correlated loss channel
    $[p_{LL},p_{LQ},p_{LN}]=[p/3,p/3,p/3]$ applied at every CZ gate. We chose the coefficient threshold $5\times10^{-3}$.
    Except for the curve annotated as non-preserving (n.p.), all runs use the protected set
    $\{Z_q\}$.}
\end{figure}

Since the total magnetization is conserved, the overlap remains at its initial value $m(t)=m(0)=1$ in the absence of loss and truncation errors. We enforce this
conservation \emph{exactly under truncation} using the protected set introduced in
Sec.~\ref{sec:conservation}: the single-site charge operators $\{Z_q\}$ are
exempt from truncation and the coefficient truncation
is applied only after each
(magnetization-conserving) bond gate. As \fref{fig:transport} shows, protection sees the noiseless $m(t)$ stay exactly at $1$ despite the use of coefficient truncation, whereas foregoing protection (curve n.p.) causes the conserved charge to leak and $m(t)$ to drift below $1$.

To model loss faithfully, we decompose each two-qubit interaction
$R_{XX}R_{YY}R_{ZZ}$ into single-qubit Clifford, $R_Z$ rotations, and CZ gates, where $R_{PP}(\theta)=e^{-i\theta P_iP_j/2}$ follows the half-angle convention of \eqref{eq:pauli_rotation}. We then
attach a \emph{correlated} two-qubit loss channel (Sec.~\ref{sec:correlated-loss})
to the CZ (entangling) gates only, reflecting that loss predominantly occurs during two-qubit gates.
We use $[p_{LL},p_{LQ},p_{LN}]=[p/3,p/3,p/3]$, so the total per-CZ loss probability
with both atoms present is $p_{LL}+2p_{LQ}=p$.
Without reset, no loss branches are introduced, and the survival-weighted overlap is reduced by exactly $(1-p)$ per CZ, i.e. $m(t)=(1-p)^{k(t)}$ for the $k(t)$ CZ gates applied up to time $t$.
Since the protected set prevents truncation from changing this overlap, the lossy lines in \fref{fig:transport} reproduce these exact scalings.
As $p$ grows (here $0.25\%$ and $0.5\%$), the overlap decays faster, reproducing the analytical scaling.

\begin{table*}[!t]
\centering
\small
\setlength{\tabcolsep}{4pt}
\begin{tabular}{@{}>{\raggedright\arraybackslash}p{0.30\textwidth}llrr@{}}
\toprule
Change / hypothesis & Commit & Decision & Step speedup & Baseline speedup \\
\midrule
Baseline & \href{https://github.com/QuEraComputing/ppvm/commit/34dcdb21ee90921e7832a3ccdb1bc3e73b78bbdd}{\texttt{34dcdb2}} & Reference & --- & $1.00\times$ \\
Clone only surviving branches & \href{https://github.com/QuEraComputing/ppvm/commit/5916523571c95b4aedbc500df1bccae5a410a6d4}{\texttt{5916523}} & Accepted & $2.73\times$ & $2.73\times$ \\
Hash gathered words in bulk & \href{https://github.com/QuEraComputing/ppvm/commit/d0cc2fe88dadb6ab258c02bf090cd67a196073e5}{\texttt{d0cc2fe}} & Accepted & $1.67\times$ & $4.57\times$ \\
Precompute row masks & \href{https://github.com/QuEraComputing/ppvm/commit/26e222391aa9fc7d5a1b94eb5ffa8ca49d059c2b}{\texttt{26e2223}} & Accepted & $1.04\times$ & $4.75\times$ \\
Scalar hash without a buffer & \href{https://github.com/QuEraComputing/ppvm/commit/34b41ceba941677b3fef69a509d2d3333c05cb2f}{\texttt{34b41ce}} & Rejected & $0.60\times$ & $2.84\times$ \\
Restore SIMD hashing & \href{https://github.com/QuEraComputing/ppvm/commit/5ea2c00f0a00841197854eb6b2cb7fa4cbd31905}{\texttt{5ea2c00}} & Restored & $1.00\times$ & $4.75\times$ \\
Read storage words directly & \href{https://github.com/QuEraComputing/ppvm/commit/4be19ee3afae3273082c217ccbd56c38369df8ee}{\texttt{4be19ee}} & Accepted & $1.06\times$ & $5.04\times$ \\
\bottomrule
\end{tabular}
\caption{\label{tab:autotune-history}Accepted and rejected autotune experiments for generalized-tableau mixture construction in the \texttt{msd-noisy} workload~\cite{ppvm_autotune_pr157}. Step speedup is the construction time of the preceding accepted implementation divided by the time for the listed implementation; baseline speedup is $2620\,\mathrm{ms}$ divided by the time for the listed implementation. Rejected candidates do not update the reference for subsequent steps. Step speedups below $1\times$ indicate a regression relative to the preceding accepted implementation.}
\end{table*}

\section{Harness and performance}
\label{sec:performance}

Optimizing PPVM requires preserving the semantics of joint quantum-classical execution. In generalized-tableau simulations, PPVM executes each shot instruction by instruction, supporting non-Clifford operations, atom loss, measurement-conditioned feed-forward, and inspection of the quantum and classical states. Compiled sampling modes in tools such as Stim~\cite{gidney2021stim}, Tsim~\cite{haenel2026tsimfastuniversalsimulator}, or Clifft~\cite{chase2026clifftfastexactsimulation} can exploit advance knowledge of circuit structure, so performance comparisons must account for differences in execution features. For observable evolution, PPVM's Pauli propagation backend achieves runtimes comparable to those of dedicated tools such as PauliPropagation.jl~\cite{rudolph2025pauli} in the benchmarks reported in Appendix~\ref{app:benchmarks}. To preserve execution semantics during optimization, we use a development workflow that checks candidate changes against reference implementations and tests of the relevant physics.

\begin{figure}[!b]
    \centering
    \includegraphics[width=\columnwidth]{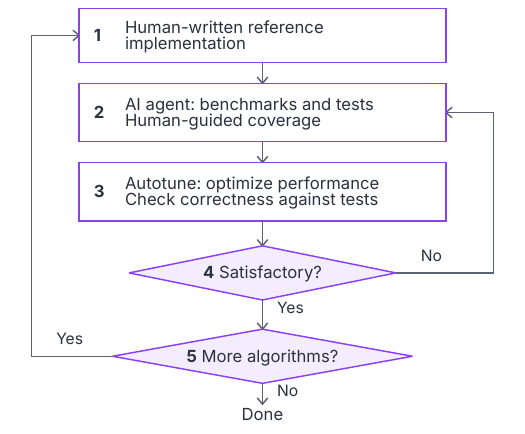}
    \caption{\label{fig:harness}
        \emph{The harness-engineering workflow used to develop PPVM.} (1) A human provides a reference implementation with autocompletion-level AI assistance. (2) An AI agent develops benchmarks and tests under human guidance. (3) The autotune evaluates candidate optimizations using benchmarks, reference comparisons, and tests. (4) If performance is unsatisfactory, steps~2--3 repeat. (5) The workflow returns to step~1 for the next algorithm, or finishes if none remain.}
\end{figure}

We developed PPVM using a harness-engineering workflow that combines human guidance with AI-assisted implementation and optimization, as shown in \fref{fig:harness}. AI agents iteratively propose and evaluate implementation changes using tests and benchmarks developed under human guidance. Detailed comparisons with other simulators are reported in Appendix~\ref{app:benchmarks}. Here, we focus on the development workflow and the evidence from recorded tuning sessions.

The workflow uses a human-written reference implementation of each algorithm and a test suite covering the relevant physics. The reference establishes the behavior to be preserved, while the tests check candidate implementations as agents explore performance optimizations. Our \emph{autotune} workflow began as a skill inspired by Andrej Karpathy's \texttt{autoresearch}~\cite{karpathy2026autoresearch,ppvm_repo}. We later developed it into an agent implementation with hardened execution logic and verification~\cite{luo2026autotune}.

Recorded autotune sessions provide concrete examples of this workflow. One session reduced MSD sampling time by $32.3\%$~\cite{ppvm_autotune_pr49}. A second session optimized the \texttt{msd-noisy} workload, which constructs a generalized-tableau mixture for magic state distillation with single-qubit depolarization and loss. It reduced mixture-construction time by a factor of $5.04$~\cite{ppvm_autotune_pr157}. Table~\ref{tab:autotune-history} summarizes the accepted and rejected experiments from this session~\cite{ppvm_autotune_pr157}.

The largest gain came from delaying tableau cloning until after branch merging and truncation. The mixture retained 2,025 branches with unchanged recorded probability statistics, while sampling time remained approximately $22\,\mu\mathrm{s}$ per shot. The session also tested whether hashing tableau words directly in a single scalar pass could outperform gathering them into a temporary buffer for SIMD hashing. On this workload, eliminating the buffer increased construction time from $552$ to $921\,\mathrm{ms}$, so the change was reverted. These accepted and rejected experiments illustrate how runtime measurements and reference checks guide optimization decisions.

\section{Conclusion}
\label{sec:conclusion}

We have introduced PPVM, an open-source framework that emulates a quantum processor together with its classical controller. This inspectable execution model provides a foundation for digital twins of quantum processors that integrate hardware noise, decoding, and adaptive control. A shared bytecode virtual machine unifies generalized tableau simulation for quantum-state sampling and measurement-conditioned feed-forward with Pauli propagation for observable evolution.

For neutral-atom hardware, we derived independent and correlated loss channels, including the altered dynamics when a gate partner is already lost, and incorporated loss and reset into both backends. We have also shown how to preserve overlap with conserved quantities under truncation and to sample approximately from truncated classical mixtures. These capabilities are demonstrated through loss-aware magic state distillation on up to $1435$ physical qubits, adaptive logical programs coupled to classical decoding, and spin-correlation dynamics in a Heisenberg model with loss.

Future work can extend PPVM toward more complete digital twins of neutral-atom hardware. One direction is a tighter integration of real-time decoding within the virtual machine, building on the existing coupling in Bloqade QEC Studio~\cite{qec-studio} and extending it to use loss reports in decoding~\cite{baranes2026loss}. Other directions include asymmetric loss in generalized tableau simulation, leakage in Pauli propagation, and measurement-conditioned classical control in backward operator evolution. Reference implementations, physics tests, and benchmarks provide a workflow for improving performance as these capabilities expand. PPVM offers an open framework for evaluating quantum protocols, hardware noise models, and classical control strategies within a shared execution environment.

\section*{Acknowledgments}

We thank Manuel S. Rudolph, Bradley A. Chase, Yue Wu, Alan Bidart, Jin-Sung Kim, and Tom Lubowe for helpful discussions. Work completed before January 2026 was carried out entirely by the human authors. From January 2026 onward, AI agents, including Claude, Cursor Composer, and OpenAI Codex, assisted with software development through the harness and agent workflows described in Sec.~\ref{sec:performance}, as well as manuscript editing and language polishing. The scientific content was developed and written by the human authors, who reviewed the revisions and remain responsible for the work.

\Urlmuskip=0mu plus 1mu\relax
\renewcommand{\bibfont}{\raggedright}
\bibliographystyle{unsrtnat}
\bibliography{ref}

\onecolumn\newpage
\appendix

\section{Kraus operators of correlated loss channels}
\label{app:correlated-loss}

We provide the Kraus operator expressions for the correlated loss channel introduced in Sec.~\ref{sec:correlated-loss}.
The channel acting on both atoms without losing either of them is described by
\begin{align}
    K_{0} = \sqrt{1 - p_{LL} - 2p_{LQ}}\, I \otimes I + \sqrt{1 - p_{LN}} \left(I \otimes L + L \otimes I\right) + L \otimes L,
\end{align}
where $I = \ket{0}\bra{0} + \ket{1}\bra{1}$ denotes the identity on the computational subspace and $L = \ket{L}\bra{L}$.

The four Kraus operators describing the loss of one atom while the other remains in the computational subspace are
\begin{align}
    K_{LQ,1} &= \sqrt{p_{LQ}}\ket{L}\bra{\alpha} \otimes I,
    \\
    K_{LQ,2} &= \sqrt{p_{LQ}}\, I \otimes \ket{L}\bra{\alpha},
\end{align}
where $\alpha\in\{0, 1\}$.
The operators $K_{LQ,1}$ and $K_{LQ,2}$ correspond to losing the first and the second atom, respectively; the state of the remaining atom is unaffected.

The four Kraus operators describing the loss of both atoms at once are
\begin{align}
    K_{LL} = \sqrt{p_{LL}}\ket{L}\bra{\alpha} \otimes \ket{L}\bra{\beta},
\end{align}
where $\alpha, \beta \in\{0, 1\}$.

Finally, the four Kraus operators describing the loss of one atom when the other atom has been lost prior to the channel are
\begin{align}
K_{LN,1} &= \sqrt{p_{LN}} \ket{L}\bra{\alpha}\otimes L,
\\
K_{LN,2} &= \sqrt{p_{LN}}\, L \otimes \ket{L}\bra{\alpha},
\end{align}
where $\alpha\in\{0, 1\}$.

\section{Benchmarks}
\label{app:benchmarks}

We compare PPVM's performance to other state-of-the-art simulation frameworks.
Unless stated otherwise, PPVM was called via its Python bindings and run on a single thread of an Apple M4 (24\,GB RAM), except for the GPU benchmarks in Sec.~\ref{app:gpu}.

\subsection{Pauli propagation}

For Pauli propagation, we pick two standard problems to benchmark:
The Transverse-Field Ising Model (TFIM),
\begin{align}
    H_{\mathrm{TFIM}} = \frac{h}{2} \sum_i X_i + \frac{J}{2} \sum_{\braket{i, j}} Z_i Z_j,
\end{align}
and the Heisenberg model introduced in Sec.~\ref{sec:transport}, \eqref{eq:heisenberg}.
Additionally, we apply depolarization noise after every qubit rotation in each Trotter step to each qubit involved in the gate.

\begin{figure}[!htbp]
    \centering
    \includegraphics[width=\columnwidth]{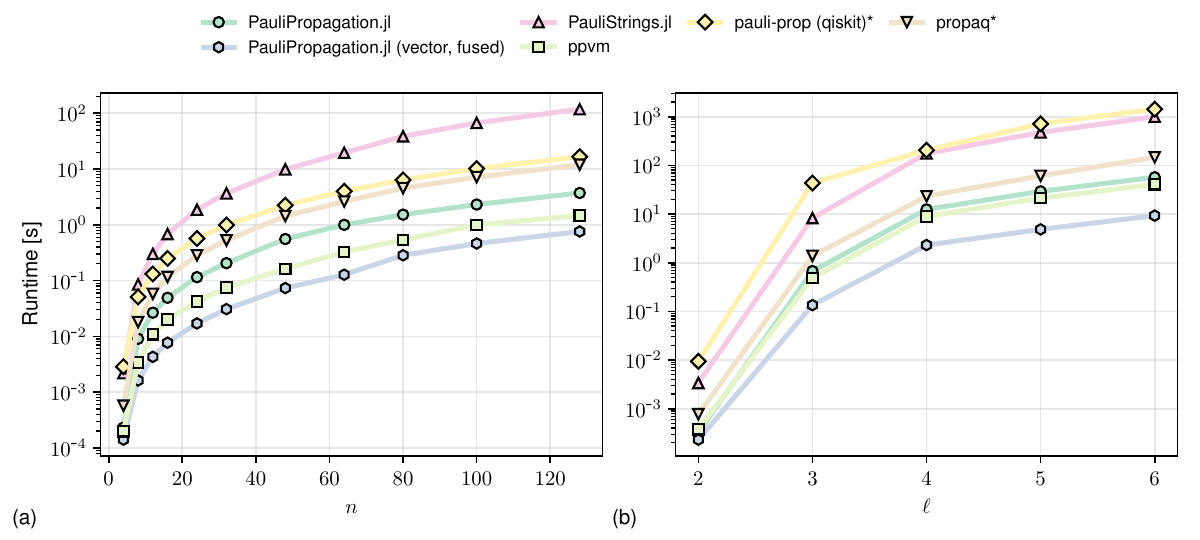}
    \caption{\label{fig:pp-benchmark}
    \emph{Benchmark comparison of PPVM with other frameworks.}
    (a) The TFIM shows that PPVM is close in performance to the vectorized runs of PauliPropagation.jl, which is the fastest.
    (b) In the Heisenberg model correlations spread quicker since each qubit has multiple neighbors, which increases the number of terms one has to track.
    All frameworks exhibit the expected polynomial scaling in size. ($\ast$) qiskit's \texttt{pauli-prop} and the \texttt{propaq} package have a distinct algorithmic difference (see main text), which is why scaling may differ.
    The parameters were $J=h$, $dt = 0.1/h$, $T = 2/h$ for the TFIM and $dt = 0.1/J$, $T = 2/J$ for the Heisenberg model (20 steps each), with a depolarization probability of $p_{\mathrm{dpl}} = 10^{-4}$ and a coefficient truncation threshold of $10^{-6}$. All benchmarks were run on a single thread of an Apple M4 processor (24\,GB RAM).
    We used PPVM (commit \texttt{731e59f}, via its Python bindings), PauliPropagation.jl~0.9.0 and PauliStrings.jl~1.10.1 on Julia~1.13.0, and \texttt{pauli-prop}~0.2.1 (Qiskit~2.4.2) and \texttt{propaq}~0.1.7 on Python~3.12.12.
    }
\end{figure}

We compare PPVM to a few other popular packages: PauliPropagation.jl~\cite{rudolph2025pauli}, which is arguably the closest package semantically;
PauliStrings.jl~\cite{loizeau2025quantum,paulistrings_v1101}, which has a broader focus but also offers all the tooling to simulate Trotter circuits efficiently;
and, finally, qiskit's \texttt{pauli-prop} package~\cite{qiskit_pauli_prop} and the recent \texttt{propaq} package~\cite{belagali2026propaq}.
Note that the latter two exhibit algorithmic differences:
in \texttt{pauli-prop} equal Pauli words are not merged after gate application, but still truncated based on the magnitude of their coefficients. Furthermore, there is no weight truncation, but the maximum number of allowed terms is limited prior to execution.
\texttt{propaq} truncates terms before merging, which also leads to slight deviations in the computed expectation values.

We also note that the reason PauliPropagation.jl is fastest is that it uses a vectorized storage rather than a map storage, which is advantageous in the considered benchmarks.

Another very recent and high-performance package is \texttt{monoprop}~\cite{miller2025simulation,miller2026monoprop}, which employs substantially different data structures to maintain the collection of Pauli strings. It was omitted from the above benchmarks as it currently does not have support for noise.

\subsection{Generalized tableau simulation}

We compare the runtime of PPVM's generalized tableau backend with other simulators that differ in their supported features and execution models (see \fref{fig:overview}(b)).

One of the most popular packages for QEC-type simulations is Stim~\cite{gidney2021stim}.
It is highly performant and significantly faster than PPVM when simulating Clifford circuits, but it does not extend beyond Clifford circuit simulations.
More recently, a number of frameworks have been developed to push beyond the Clifford and into the near-Clifford regime:
among them are Tsim~\cite{haenel2026tsimfastuniversalsimulator}, SOFT~\cite{li2025soft}, and Clifft~\cite{chase2026clifftfastexactsimulation}.
Both Tsim and Clifft pre-compile circuits leading to very high sampling rates, comparable to Stim, and beyond what PPVM's step-wise execution per shot can offer.
However, they only support very limited control flow, whereas PPVM supports arbitrary classical control flow.
Finally, PPVM supports instruction-by-instruction execution and state inspection, which the compilation-based algorithms used by Tsim and Clifft do not support.
SOFT similarly uses generalized tableau simulation, with GPU parallelization across independent shots.

\begin{figure}[!htbp]
    \centering
    \includegraphics[width=\columnwidth]{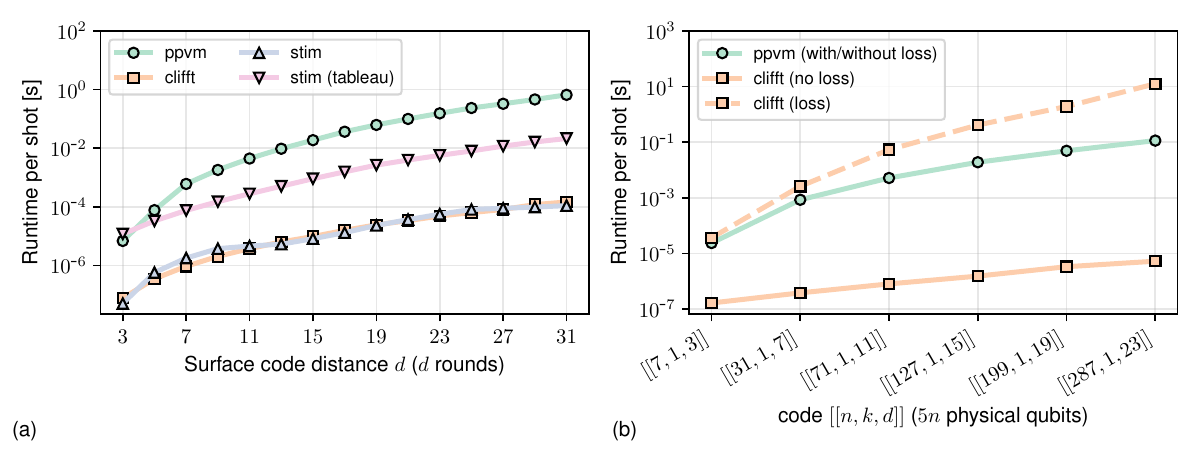}
    \caption{\label{fig:tableau-loss-benchmark}
    \emph{Benchmark comparison of PPVM with other frameworks.}
    \textbf{(a)} We compare PPVM with Clifft and Stim when running the rotated surface code experiment as a function of code distance $d$ for rounds $r=d$~\cite{gidney2021stim}.
    As we do not include loss or any non-Clifford noise or gates, these circuits can be run by Stim too.
    The compiled sampling in Stim and Clifft makes them both much more performant.
    We also compare against using the Stim tableau simulator, which is also faster.
    \textbf{(b)} We run the same magic state distillation circuits as in \fref{fig:msd-runtime}, with and without loss.
    Without loss, the compilation makes Clifft orders of magnitude more efficient than PPVM.
    However, the repeated execution forces Clifft to recompile every shot, which is why there is a significant runtime penalty when simulating loss in Clifft.
    In contrast, simulating loss in PPVM is essentially free.
    Note that we replaced every correlated loss channel by an independent loss channel, as Clifft does not support correlated loss yet.
    All benchmarks were run on a single thread of an Apple M4 processor (24\,GB RAM), with PPVM called via its Python bindings.
    We used Stim v1.16.0 and Clifft v0.10.1.
    }
\end{figure}

Atom loss affects the relative performance of the simulators in the near-Clifford benchmarks considered here.
As shown in \fref{fig:tableau-loss-benchmark}, where Clifft is far more efficient than PPVM without including loss.
With loss, however, PPVM is more performant.
We do note that Clifft's loss simulation capabilities are a recent feature and we expect performance to improve in the near future.

\subsection{GPU acceleration}
\label{app:gpu}

\begin{figure}[!htbp]
    \centering
    \includegraphics[width=\columnwidth]{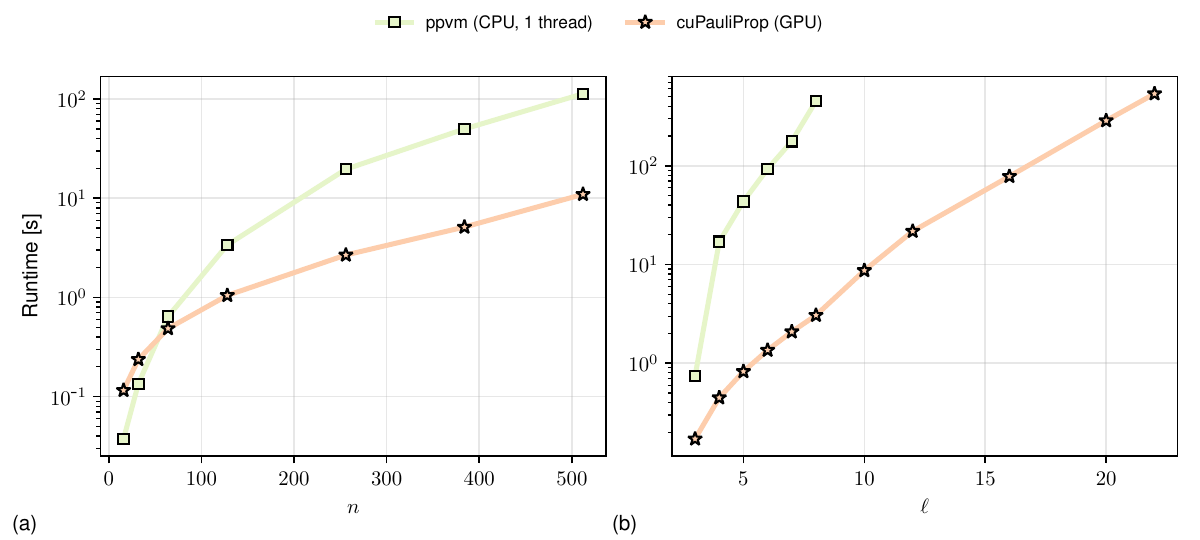}
    \caption{\label{fig:gpu-benchmark}
    \emph{PPVM's CPU backend compared to its GPU backend based on cuPauliProp.}
    (a) TFIM and (b) Heisenberg model with the same parameters as in \fref{fig:pp-benchmark}.
    The CPU runs are restricted to $\ell\le 8$ for the Heisenberg model due to their runtime.
    Both backends were run on an NVIDIA GH200 Grace Hopper Superchip, the CPU backend on a single thread of the Grace CPU as a runtime reference, the GPU backend on the full Hopper GPU.
    Runtimes are minima over three samples.
    We used PPVM (commit \texttt{9bc0bcb}) and cuPauliProp~0.5.0.
    }
\end{figure}

Pauli propagation maps well onto GPUs: every gate acts independently on each term of the Pauli sum.
We therefore added an experimental GPU backend to PPVM that uses NVIDIA's cuPauliProp library~\cite{cupauliprop}.
The backend exposes the same gate interface as PPVM's CPU Pauli sum.
\fref{fig:gpu-benchmark} compares the two backends on the benchmarks of \fref{fig:pp-benchmark}.
Both backends agree on the expectation value to within $10^{-11}$ and on the number of remaining terms exactly at every system size.

The single-threaded CPU runs are included as a reference for the runtime rather than as a competitive comparison.
With this in mind, the GPU is slower for small Pauli sums, as each gate application carries a fixed overhead.
For the TFIM, it overtakes the single CPU thread at $n=64$ (about $10^5$ terms), for the Heisenberg model already at $\ell=3$.
The runtime ratio reaches $10\times$ for the TFIM at $n=512$ and $148\times$ for the Heisenberg model at $\ell=8$, where the number of terms grows much faster with system size.
On the GPU, we simulated the Heisenberg model up to $\ell=22$ ($484$ qubits, $1.5\times10^7$ terms) in about nine minutes.

\end{document}